\documentclass[conference]{IEEEtran}
\IEEEoverridecommandlockouts
\usepackage{cite}
\usepackage{amsmath,amssymb,amsfonts}
\usepackage{algorithmic}
\usepackage{graphicx}
\usepackage{subcaption}
\usepackage{textcomp}
\usepackage{xcolor}
\usepackage{bm}
\usepackage{xspace}
\usepackage[labelformat=simple]{subcaption}

\def\BibTeX{{\rm B\kern-.05em{\sc i\kern-.025em b}\kern-.08em
    T\kern-.1667em\lower.7ex\hbox{E}\kern-.125emX}}
\begin{document}
\newcommand{\fig}{Fig.\xspace}
\newcommand{\cm}[1]{}

\textwidth 7.5in
\oddsidemargin -0.5in
\topmargin -1 in
\textheight 10.1 in

\title{A New Paradigm for Device-free Indoor Localization: Deep Learning with Error Vector Spectrum in Wi-Fi Systems}
\author{\IEEEauthorblockN{Wen Liu, An-Hung Hsiao, Li-Hsiang Shen and Kai-Ten Feng}
\IEEEauthorblockA{Department of Electronics and Electrical Engineering, National Yang Ming Chiao Tung University, Hsinchu, Taiwan\\
Email: swimmingliu880208.ee10@nycu.edu.tw, e.c@nycu.edu.tw, gp3xu4vu6.cm04g@nctu.edu.tw, and ktfeng@nycu.edu.tw
}}
\maketitle

\begin{abstract}
The demand for device-free indoor localization using commercial Wi-Fi devices has rapidly increased in various fields due to its convenience and versatile applications. However, random frequency offset (RFO) in wireless channels poses challenges to the accuracy of indoor localization when using fluctuating channel state information (CSI). To mitigate the RFO problem, an error vector spectrum (EVS) is conceived thanks to its higher resolution of signal and robustness to RFO. To address these challenges, this paper proposed a novel error vector assisted learning (EVAL) for device-free indoor localization. The proposed EVAL scheme employs deep neural networks to classify the location of a person in the indoor environment by extracting ample channel features from the physical layer signals. We conducted realistic experiments based on OpenWiFi project to extract both EVS and CSI to examine the performance of different device-free localization techniques. Experimental results show that our proposed EVAL scheme outperforms conventional machine learning methods and benchmarks utilizing either CSI amplitude or phase information. Compared to most existing CSI-based localization schemes, a new paradigm with higher positioning accuracy by adopting EVS is revealed by our proposed EVAL system.
\end{abstract}

\section{Introduction}
Smart home appliances have sprung up one after another in this era of the rise of the Internet of Things (IoT). Device-free indoor localization has emerged as a crucial research topic in recent years, given its potential to revolutionize how we interact with indoor environments. An accurate indoor location system enables a variety of location-based services, including intrusion detection, intelligent building management, and targeted advertising. Nowadays, camera-based detection \cite{camera} and infrared sensors \cite{infrared} are the most common techniques for device-free indoor localization. However, these methods have several limitations that restrict their practical application. Camera-based detection, for example, raises concerns regarding personal privacy and is susceptible to blind spot problems. These limitations have led to increased interest in alternative methods for device-free indoor localization. Furthermore, infrared sensor-based schemes are only applicable when the individual is within the line-of-sight (LoS), making them ineffective when the person is stationary or in a non-line-of-sight (NLoS) area. To address this challenge, researchers have explored alternative technologies that utilize wireless signals to estimate the position of humans indoors \cite{acm}.

The received signal strength indicator (RSSI) is a popular and straightforward approach for indoor localization, as indicated in \cite{RSSI_HOI} and \cite{RSSI_trilateration}. This technique estimates the distance between a transmitter and a receiver by analyzing the strength of the received signal. By measuring the RSSI of wireless signals emitted by Wi-Fi access points, Bluetooth beacons, or other wireless devices, it is possible to obtain the location of individuals in indoor environments. On the other hand, time-of-flight (ToF) technology can also be utilized for device-free indoor localization by measuring the time it takes for a signal to travel from a transmitter to a receiver. \cite{UWB_TDoA} investigates the time difference of arrival, and \cite{UWB_TWTF} exploits two-way ToF to calculate the range measurements. However, either the RSSI-based or ToF-based localization method is prone to signal attenuation, where the accuracy decreases with increasing distance between the access points (APs) or beacons. Additionally, they are susceptible to multipath interference and environmental variations, which can significantly affect the accuracy of the method. Moreover, RSSI needs to achieve high-precision measurement by deploying multiple APs or beacons, and the receiver required for ToF is quite expensive.

In order to solve the estimation error from multipath effect and signal attenuation, channel state information (CSI) is an advanced approach for indoor localization that utilizes the subtle changes in the wireless signal to estimate the position of the target, which is divided into amplitude and phase information. In \cite{CSI_Amp_RF, CSI_Amp_SVM, CSI_Amp}, they adopt learning-based methods with CSI amplitude as input to classify the position of a human. In \cite{csi1, csi2}, they adopt deep learning based methods leveraging CSI information to detect human presence in indoor environments. However, the CSI amplitude signal characteristics become unresponsive when the person is in NLoS positions, which is pointed out in \cite{CSI_NLoS}. In contrast, the CSI phase information is susceptible to the indoor environment, which is capable of detecting slight fluctuation due to reflecting signals. Therefore, papers of \cite{CSI_Phs_NB, CSI_Phs_KNN} have utilized CSI phase information as a characteristic signal to the traditional machine learning algorithm. Nonetheless, the raw CSI phase information substantially suffers from the random frequency offset (RFO) problem, which leads to indescribability. This requires specialized hardware as well as sophisticated signal-processing algorithms, limiting its practical applicability.

For the above-mentioned reasons, we have alternatively employed a unique feature of wireless signals, i.e., error vector spectrum (EVS) to accurately estimate the position of humans in indoor environments. EVS is capable of surmounting the low localization resolution in NLoS to break through the such predicament. We achieve this by analyzing the error vector of the received wireless signals, which enables us to identify the exact location of the target. Moreover, due to several emerging advanced sensing specifications and techniques applied in wireless local area networks (WLANs) for IEEE 802.11ax/be/bf \cite{ax,be,bf} and even to-be-discussed next-generation Wi-Fi 8 \cite{wifi8}, where most of them are restricted by certain hardware limitations, which provokes a difficulty of development. Therefore, we require open equipment, e.g., OpenWiFi \cite{openwifi} to develop advanced sensing and localization schemes without laborious switching platforms.

 The main contributions of this paper are listed as follows.
\begin{itemize}
    \item Best to our knowledge, we are the first work to propose an error vector assisted learning (EVAL) using OpenWiFi system to improve the accuracy of device-free indoor localization, in which the data preprocessing and error vector based classifier are presented to utilize deep neural network learning. Our proposed method outperforms existing approaches in terms of accuracy.
    \item To overcome the limitations of CSI caused by the RFO issue, our proposed system designs are based on the employment of EVS as a potential feature extracted from the physical layer signal. EVS carries diversified information, which can provide a more accurate estimation of the indoor location compared to conventional CSI-based methods.
    \item Experiment trials are conducted by establishing an OpenWiFi-based platform, which is compatible with existing WLAN systems. We have demonstrated the advantages and effectiveness of the proposed EVAL system for indoor localization. The results have demonstrated that EVAL outperforms other learning algorithms and mechanisms utilizing CSI information, which achieves the highest accuracy in device-free localization.    
\end{itemize}

\section{System Architecture}

\begin{figure}
  \centering
  \includegraphics[width=0.45\textwidth]{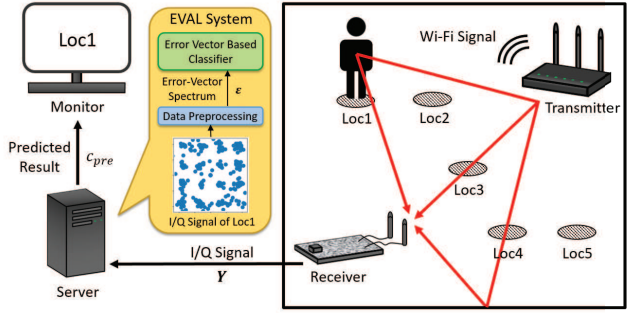}
  \caption{System architecture for indoor localization by deploying commercial Wi-Fi devices.}
  \label{fig:system}
\end{figure}

 Fig.~\ref{fig:system} illustrates the system architecture of our proposed approach for indoor localization in a single-room environment. We set up commercial off-the-shelf Wi-Fi APs as the transmitter and an open-source software-defined radio (SDR) implementation of a Wi-Fi AP as the receiver. The Wi-Fi signal packets are broadcast from the transmitter and collected by the receiver. We employ the orthogonal frequency division multiplexing (OFDM) technique in IEEE 802.11ax/be \cite{ax,be} for acquiring diversified channel characteristics in the indoor environment. These OFDM-partitioned subcarriers are orthogonal to each other within a channel, ensuring that they cause no interference to each other. The received signal can be expressed as
\begin{equation}
    \label{eq:channel_model}
    \bm{Y}=\bm{H}\bm{X}+\bm{W}, 
\end{equation}
where $\bm{Y}$ and $\bm{X}$ represent the received and transmitted complex signal matrices of dimension $K\times N$, respectively, whilst $\bm{W}$ accounts for additive white Gaussian noise (AWGN) with dimension $K\times N$. Here, $K$ and $N$ are the maximum numbers of subcarriers and symbols, respectively. $\bm{H}$ is the complex channel response diagonal matrix of dimension $K\times K$ and can be further represented as
\begin{equation}
    \label{eq:channel_matrix}
    \bm{H}={\rm \bm{diag}}\left(h_1,h_2,...,h_k,...,h_K\right), 
\end{equation}
where $k\in [1,...,K]$ indicates the subcarrier index. Each element of $\bm{H}$, which is $h_k$, is generally described by the channel impulse response (CIR) to the multipath effect of the channel, which also characterizes the multipath propagation attenuated, scattered, and faded in respective subcarriers. Thus, each element of $\bm{H}$ on the diagonal is modeled as
\begin{equation}
    \label{eq:channel_element}
    h_k=\sum_{m=1}^{M} \alpha_{m,k}e^{-j2\pi\frac{d_m}{\lambda_k}}=|h_k |e^{j(\angle{h_k})}, 
\end{equation}
where $M$ is the total number of propagation multipath received at the receiver end. $\lambda$ represents the wavelength, $\alpha_{m,k}$ stands for complex signal attenuation on the $m^{th}$ path at $k^{th}$ subcarrier, and $d_m$ refers to the propagation length on the $m^{th}$ path. Joint factors of $\alpha_{m,k}$ and $d_m$ reflecting channel variation can be obtained from the amplitude $\left|h_k \right|$ and phase $\angle{h_k}$ of the $k^{th}$ subcarrier. As depicted in Fig.~\ref{fig:system}, the received signal is then processed to extract the in-phase and quadrature (I/Q) signal information, where the I/Q signal after equalization is displayed as the constellation diagram. The data preprocessing block transforms I/Q signals into EVS $\pmb{\varepsilon}$ expressed as
\begin{equation}
    \label{eq:evs_vector}
    \pmb{\varepsilon}=\left[\varepsilon_1,\varepsilon_2,...,\varepsilon_k,...,\varepsilon_K\right]. 
\end{equation}
Each element of $\pmb{\varepsilon}$ can be expressed as the amplitude and phase combination form
\begin{equation}
    \label{eq:evs_amphs}
    \varepsilon_k=|\varepsilon_k|e^{j(\angle{\varepsilon_k})}, 
\end{equation}
where $|\varepsilon_k|$ and $\angle{\varepsilon_k}$ represent the amplitude and phase information of EVS and can be utilized as the localization features, respectively.

\section{Proposed Error Vector Assisted Learning (EVAL) System}

\begin{figure}
  \centering
  \includegraphics[width=0.47\textwidth]{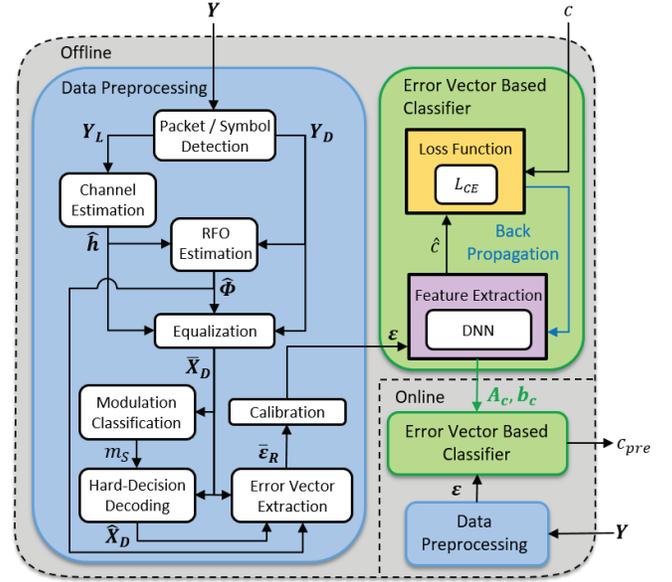}
  \caption{Schematic diagram of the proposed EVAL system.}
  \label{fig:block}
\end{figure}

The EVAL system for indoor localization consists of offline and online stages, as shown in Fig.~\ref{fig:block}. In the offline stage, the received I/Q signal undergoes data preprocessing and is transformed into EVS. Feature extraction is performed using a deep neural network (DNN) with the EVS as input. The weights and biases are obtained by training the error vector based classifier with the help of a loss function and the backpropagation mechanism. During online session, the received I/Q signal is preprocessed in the same manner as in the offline section and then transformed into EVS. The EVS is then fed into the well-trained DNN model, and the real-time localization results can be obtained. Detailed information about EVAL is elaborated as follows.

\begin{figure}
  \centering
  \includegraphics[width=0.47\textwidth]{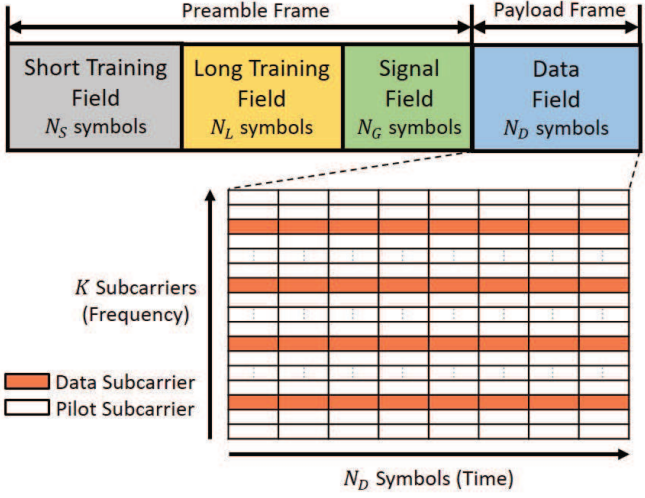}
  \caption{The OFDM frame structure in IEEE 802.11 series.}
  \label{fig:frame}
\end{figure}

\subsection{Data Sampling and Disintegration}

To extract meaningful information from I/Q signals, we have to first detect packets and symbols. The frame structure of OFDM in 802.11ax/be, depicted in Fig.~\ref{fig:frame}, is composed of a preamble and a payload frame. The preamble consists of the short training field (STF), long training field (LTF), and signal field (SF), while the payload frame contains only the data field (DF). Therefore, to convert the received I/Q signal from a bit stream to a packet form and decompose it into a frame, packet, and symbol detection are required. Thus, the received matrix can be expressed as
\begin{equation}
    \label{eq:frame_structure}
    \bm{Y} = \left[\bm{Y}_S,\bm{Y}_L,\bm{Y}_G,\bm{Y}_D\right], 
\end{equation}
where $\bm{Y}_S$, $\bm{Y}_L$, $\bm{Y}_G$, and $\bm{Y}_D$ are STF, LTF, SF, and DF matrix, with the respective dimension of $K\times N_S$, $K\times N_L$, $K\times N_G$, and $K\times N_D$. However, the STF $\bm{Y}_S$ only utilizes only one-fourth of the total subcarriers, resulting in low channel resolution. On the other hand, the SF $\bm{Y}_G$ is used to describe the basic information of the packet, such as MCS and length of the packet, but it always uses binary phase-shift keying (BPSK) modulation. Therefore, we apply LTF $\bm{Y}_L$ and DF $\bm{Y}_D$ for the following data preprocessing procedure. As for the sampling in an actual receiver, the received signal is down-converted to the baseband by a local oscillator of the receiver with local carrier frequency, which leads to phase rotation due to carrier frequency mismatch between the AP transmitter and receiver. Given the effect of frequency offset and \eqref{eq:channel_model}, the received signal after down-converted sampling and symbol detection can be obtained as
\begin{align}
    \label{eq:received_ltf_element}
    y_{L_{k,n}}&=
    {e^{-j\phi_n}}{h_k}{x_{L_{k,n}}+w_{k,n},\quad \forall{n}=1,2,...,N_L}, \\
    \label{eq:received_df_element}
    y_{D_{k,n}}&=
    {e^{-j\phi_n}}{h_k}{x_{D_{k,n}}+w_{k,n},\quad \forall{n}=1,2,...,N_D}, 
\end{align}
where $x_{L_{k,n}}$, $y_{L_{k,n}}$, $x_{D_{k,n}}$, and $y_{D_{k,n}}$ correspond to the transmit and receive LTF and DF I/Q symbols at the $k^{th}$ subcarrier and $n^{th}$ symbol. Noted that the term $e^{-\phi_n}$ represents the RFO on the $n^{th}$ symbol and this term is identical among all subcarriers of each OFDM symbol. 

\subsection{Channel and Frequency Offset Estimation}
The LTF symbol is used for fine-grained timing synchronization and channel estimation, in which the estimated channel response can be modeled as
\begin{equation}
    \label{eq:csi_vector}
    \hat{\bm{h}}=[\hat{h}_1,\hat{h}_2,...,\hat{h}_k,...,\hat{h}_K], 
\end{equation}
where $\hat{\bm{h}}$ is known as CSI vector. The AP transmits $N_L$ of the same LTF symbols with BPSK to estimate the channel. In other words, the transmitted LTF $x_{L_{k,n}}$ is known. Thus, we can conduct a generic sample average for channel estimation for each element in $\hat{\bm{h}}$ by
\begin{equation}
    \label{eq:csi_element_sub1}
    \hat{h}_k=\frac{1}{N_L}\sum_{n=1}^{N_L} y_{L_{k,n}} x_{L_{k,n}}. 
\end{equation}
According to \eqref{eq:received_ltf_element}, we can rewrite \eqref{eq:csi_element_sub1} as
\begin{align}
    \hat{h}_k&=
    \frac{1}{N_L}\sum_{n=1}^{N_L} 
    \left({e^{-j\phi_n}}{h_k}{x_{L_{k,n}}}+w_{k,n}\right) x_{L_{k,n}} \notag\\
    \label{eq:csi_element_sub2}
    &=\frac{1}{N_L}\sum_{n=1}^{N_L} 
    \left({e^{-j\phi_n}}{h_k}+w_{k,n}{x_{L_{k,n}}}\right), 
\end{align}
where $x_{L_{k,n}}$ can be removed in the first equation due to LTS being transmitted under BPSK and $x_{L_{k,n}} \in [-1,1]$. Substituting \eqref{eq:channel_element} into \eqref{eq:csi_element_sub2} yields
\begin{equation}
    \label{eq:csi_element_sub3}
    \hat{h}_k
    =\frac{1}{N_L}\sum_{n=1}^{N_L} 
    \sum_{m=1}^{M} \left(\alpha_{m,k}e^{-j\left(2\pi\frac{d_m}{\lambda_k}+\phi_n\right)}
    +w_{k,n}{x_{L_{k,n}}}\right). 
\end{equation}
Since LTF is with a short time of length 8 $\mu$sec in the specification, we can ignore the difference between each LTF symbol and AWGN $w_{k,n}$. Eventually, \eqref{eq:csi_element_sub3} can be simplified as
\begin{equation}
    \label{eq:csi_element}
    \hat{h}_k\approx
    e^{-j\phi_n}\sum_{m=1}^{M}\alpha_{m,k}e^{-j2\pi\frac{d_m}{\lambda_k}}
    =\left|\hat{h}_k \right|e^{j(\angle{\hat{h}_k})}. 
\end{equation}
So far, we have derived the formula of CSI completely and rigorously. Note that the RFO term $\phi_n$ can cause fortuitous changes in the CSI phase information $\angle{\hat{\bm{h}}}$, making it unpredictable. To solve the RFO problem, the payload frame is inserted with the pilot subcarrier which is shown in Fig.~\ref{fig:frame}. The pilot subcarrier is used to track the phase variations and remove the RFO of the corresponding DF symbol. After extracting the pilot subcarriers from each symbol of DF, we define the estimated RFO matrix $\hat{\bm{\Phi}}$ with a dimension of $N_D\times N_D$, which is a diagonal matrix as
\begin{equation}
    \label{eq:rfo_matrix}
    \hat{\bm{\Phi}}={\rm \bm{diag}}\left(\hat{\phi}_1,\hat{\phi}_2,...,\hat{\phi}_n,...,\hat{\phi}_{N_D}\right)
\end{equation}
with each element obtained as \cite{openOFDM}
$\hat{\bm{\Phi}}_{nn}\!=\!\hat{\phi}_n\!=\!\angle{ \left(\sum_{\psi\in{\Psi}} y_{D_{\psi,n}} \hat{h}_\psi\right)}$where $\psi$ refers to the subcarrier index of the pilot subcarrier. $\Psi$ is the pilot subcarrier set, which is fixed for each OFDM symbol under the 802.11a WLAN standard. Therefore, we can obtain the estimated RFO matrix $\hat{\bm{\Phi}}$ using the CSI $\hat{h}$ and the received pilot subcarrier $y_{D_{\psi,n}}$.

\subsection{Equalization}
We implement a zero-forcing (ZF) equalizer to eliminate the channel response in the received DF matrix $\bm{Y}_D$ with the aid of CSI $\hat{h}$. We also apply the estimated RFO matrix $\hat{\bm{\Phi}}$ to eradicate phase rotation. We can then obtain the equalized symbol matrix $\bar{\bm{X}}_{D}$ with dimension of $K\times N_D$ as
\begin{equation}
    \label{eq:equalization_matrix}
    \bar{\bm{X}}_D=\bm{Q} \bm{Y}_D e^{-j\hat{\bm{\Phi}}}, 
\end{equation}
where $\bm{Q}$ represents the ZF equalizer diagonal matrix with the dimension $K\times K$. Each diagonal element of $\bm{Q}$ is the reciprocal of $\hat{h}$, which can be expressed as
\begin{equation}
    \label{eq:equalizer}
    \bm{Q}_{kk}=q_k={\hat{h}_k}^{-1}.
\end{equation}
Based on \eqref{eq:received_df_element} and \eqref{eq:equalizer}, we can formulate each element of $\bar{\bm{X}}_{\bm{D}}$ as
\begin{align} \label{eq:equalization_element}
    \bar{x}_{D_{k,n}}&=e^{-j\hat{\phi}_n}{q_k}{y_{D_{k,n}}} =e^{-j\hat{\phi}_n}{\hat{h}_k}^{-1}(e^{-j\phi_n}{h_k}{x_{D_{k,n}}}+w_{k,n}). 
\end{align}
By employing the ZF equalizer and estimated RFO matrix, we have successfully obtained the equalized symbol matrix.

\begin{figure}
  \centering
  \begin{subfigure}[b]{0.24\textwidth}
    \includegraphics[width=\textwidth]{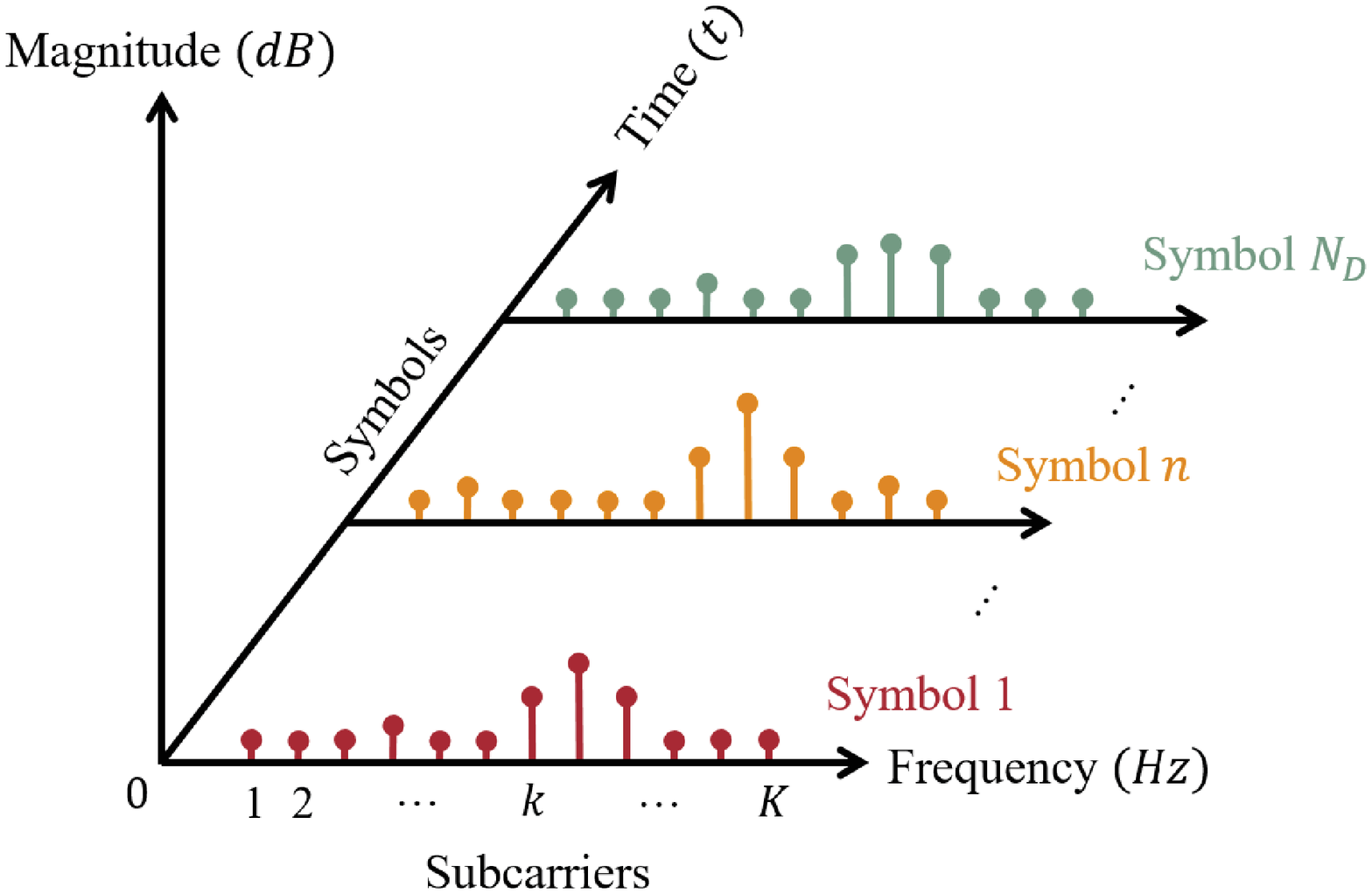}
    \caption{\footnotesize}
    \label{fig:evs_a}
  \end{subfigure}
  \begin{subfigure}[b]{0.24\textwidth}
    \includegraphics[width=\textwidth]{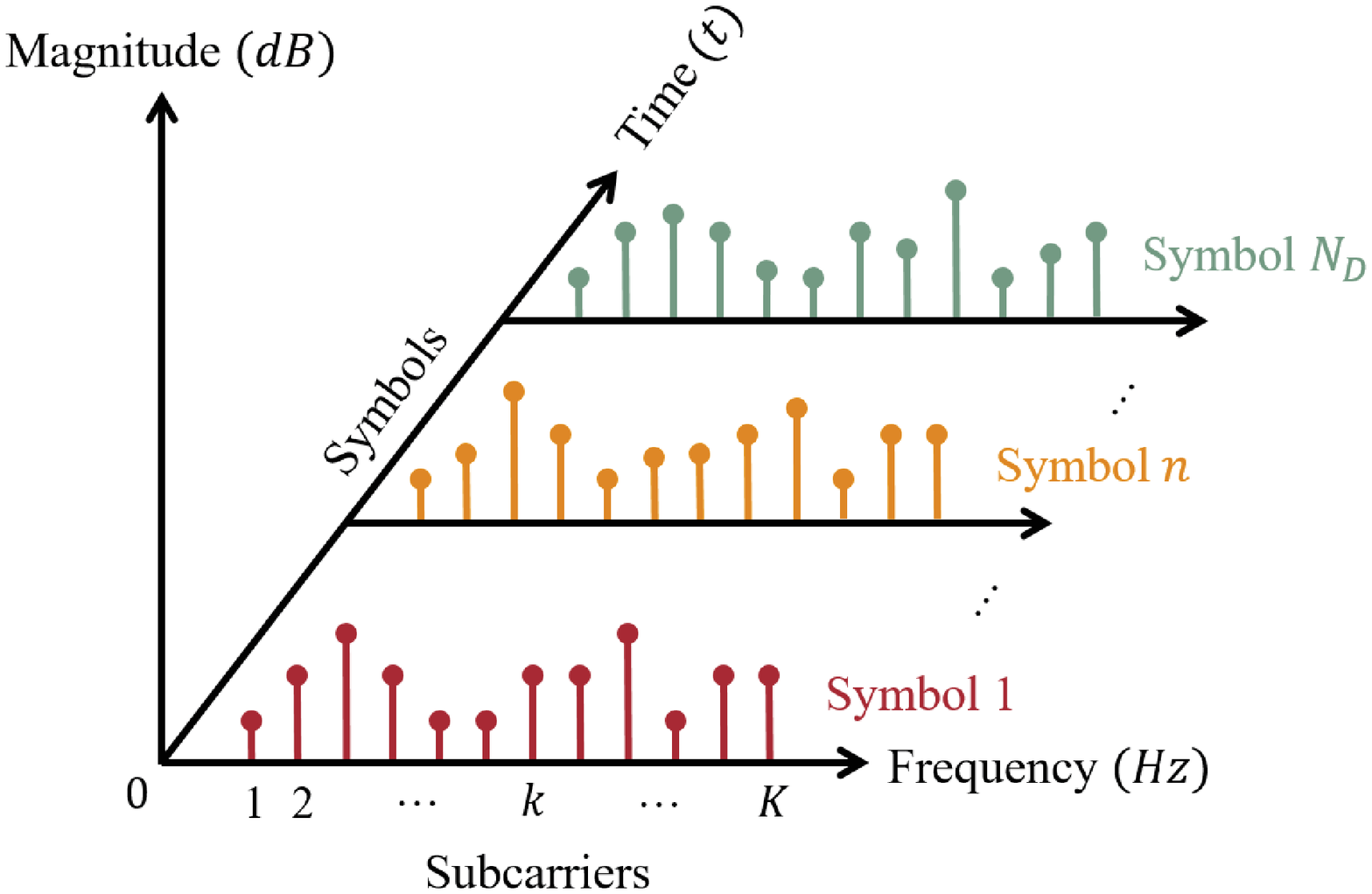}
    \caption{\footnotesize}
    \label{fig:evs_b}
  \end{subfigure}
  \\
  \begin{subfigure}[b]{0.24\textwidth}
    \includegraphics[width=\textwidth]{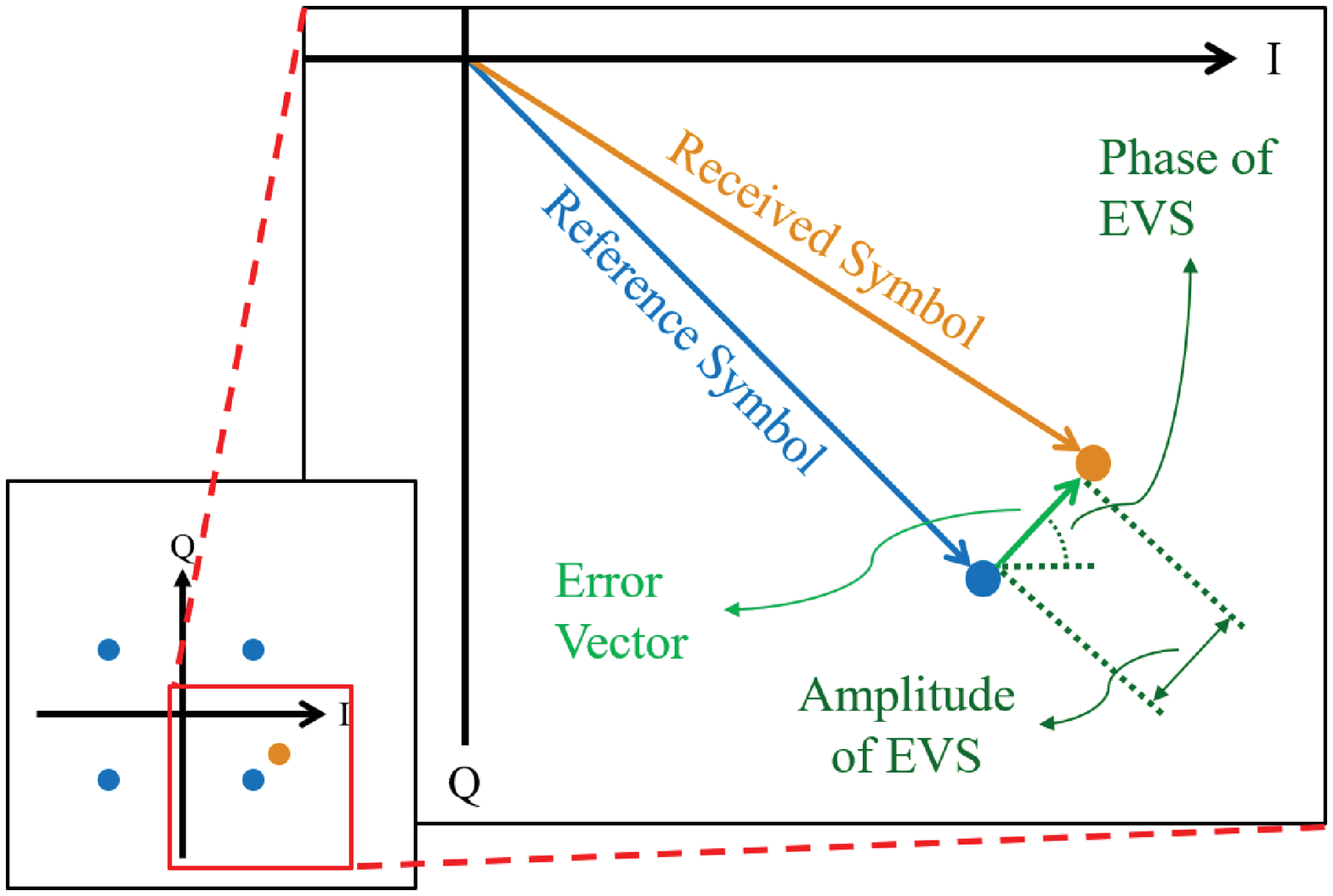}
    \caption{\footnotesize}
    \label{fig:evs_c}
  \end{subfigure}
  \begin{subfigure}[b]{0.24\textwidth}
    \includegraphics[width=\textwidth]{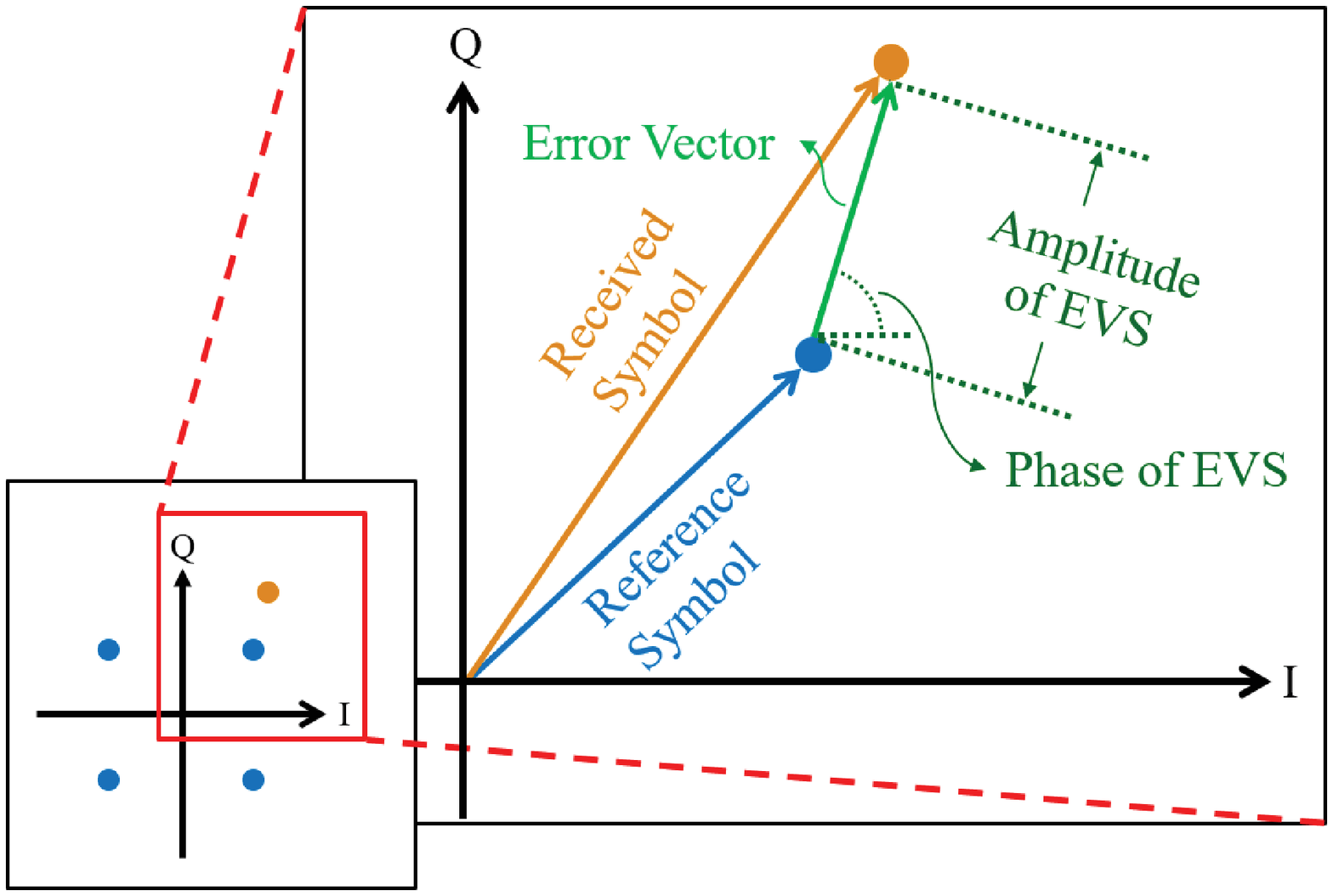}
    \caption{\footnotesize}
    \label{fig:evs_d}
  \end{subfigure}
\caption{The spectrum variation of OFDM symbol in time diagram for a stationary person at (a) location $i$ and (b) location $j$. The schematic diagrams of the EVS are respectively shown at (c) location $i$ and (d) location $j$.}
\label{fig:evs}
\end{figure}

\subsection{Error Vector Spectrum Modeling}
In Fig.~\ref{fig:evs}, the spectrum variation of OFDM symbols is shown for two cases, i.e., when a human remains stationary at locations $i$ and $j$. The corresponding cases are depicted in Figs.~\ref{fig:evs_a} and ~\ref{fig:evs_b}, respectively. The $x$-axis, $y$-axis, and $z$-axis represent the frequency, time, and degree of channel distortion, respectively. As can be seen in the figure, the same subcarrier experiences similar channel attenuation when propagating at the same frequency under the same case. Additionally, different cases can be distinguished from each other due to the channel environment diversity caused by human blocking in different locations. Based on this feature, we can recover the distortion of each subcarrier and obtain the channel characteristics. Figs.~\ref{fig:evs_c} and \ref{fig:evs_d} show schematic diagrams corresponding to the above cases. We define the Euclidean distance and phase rotation degree between the equalized and reference symbols as the amplitude and phase of the EVS, respectively. However, each symbol may be transmitted from a different constellation point after modulation. Therefore, we need to utilize modulation classification and hard-decision decoding to obtain the reference symbol corresponding to the received symbol. 802.11 series WLAN adopts four types of modulation schemes: BPSK, quadrature phase-shift keying (QPSK), 16-quadrature amplitude modulation (16-QAM), and 64-QAM. To classify the modulation methods for each equalized symbol matrix $\bar{\bm{X}}_D$, we deploy a k-means clustering algorithm \cite{k-means}. Therefore, the modulation scheme $m_S$ can be classified by the transfer function $f_K(\cdot)$ as
\begin{equation}
    \label{eq:modulation_classification}
    m_S=f_K(\bar{\bm{X}}_{\bm{D}}), 
\end{equation}
where $m_S \in \{2, 4, 16, 64\}$ is the modulation order determined by clustering transfer function $f_K$, where ${\mathbb{C}}^{K\times N_D} \rightarrow {\mathbb{N}}^{1\times 1}$. According to the adopted modulation, we employ hard-decision decoding to recover the equalized symbol to the reference symbol with the decoding transfer function $f_D(\cdot)$ as
\begin{equation}
    \label{eq:hard_decision_decoding}
    \hat{\bm{X}}_{{D}}=f_D(\bar{\bm{X}}_{\bm{D}},m_S), 
\end{equation}
where $\hat{\bm{X}}_{{D}}$ is the estimated reference symbol matrix with dimension of $K\times N_D$. Furthermore, both $\hat{\bm{X}}_{{D}}$ and $\bar{\bm{X}}_{{D}}$ are complex with dimension of ${\mathbb{C}}^{K\times N_D}$. To eliminate the redundant estimated RFO term from \eqref{eq:equalization_element}, the raw EVS $\bm{E_R}$ with dimension $K\times N_D$ can be derived as
\begin{equation}
    \label{eq:evs_matrix}
    \bm{E}_R=\bar{\bm{X}}_{{D}} - e^{-j\hat{\bm{\Phi}}} \hat{\bm{X}}_{{D}}.
\end{equation}
According to \eqref{eq:csi_element}, \eqref{eq:rfo_matrix}, and \eqref{eq:equalization_element}, each element of $\bm{E}_R$ is derived as
\begin{align}
    \varepsilon_{R_{k,n}}
    &=e^{-j\hat{\phi}_n}{\hat{h}_k}^{-1}(e^{-j\phi_n}{h_k}{x_{D_{k,n}}}+w_{k,n})
    -e^{-j\hat{\phi}_n}{\hat{x}_{D_{k,n}}} \notag \\
    &=e^{-j\hat{\phi}_n}{\hat{h}_k}^{-1}w_{k,n}.
\end{align}
Once the reference symbol $x_{D_{k,n}}$ has been eliminated by the estimated reference symbol $\hat{x}_{D_{k,n}}$, we can proceed to derive the signal based on equation \eqref{eq:csi_element_sub2} as
\begin{align}
    \varepsilon_{R_{k,n}}
    &=e^{-j\hat{\phi}_n}
    {\left(e^{-j\phi_n}\sum_{m=1}^M\alpha_{m,k}e^{-j2\pi\frac{d_m}{\lambda_k}}\right)}^{-1}
    w_{k,n} \nonumber \\
    \label{eq:evs_element}
    &={\left(\sum_{m=1}^M\alpha_{m,k}e^{-j2\pi\frac{d_m}{\lambda_k}}\right)}^{-1}w_{k,n}.
\end{align}
The above equation demonstrates that the error vector can eliminate the RFO $\phi_n$, making the EVS phase information more stable and reliable compared to the CSI phase information when used for indoor localization. To transform the raw EVS matrix into vector form, we take the mean of the symbols, which is a similar processing method to the sample average method used for CSI. Accordingly, the raw EVS can be expressed as
\begin{equation}
    \label{eq:raw_evs_vector}
    \bar{\pmb{\varepsilon}}_{{R}}=[\bar{\varepsilon}_{R_1},\bar{\varepsilon}_{R_2},...,
    \bar{\varepsilon}_{R_k},...,\bar{\varepsilon}_{R_K}],
\end{equation}
where $\bar{\varepsilon}_{R_k}=\frac{1}{N_D}\sum_{n=1}^{N_D}\varepsilon_{R_{k,n}}.$
Generally, the $N_D$ is larger than $N_L$, implying that there exist more symbols available for estimating EVS than CSI. As a result, the EVS has the potential to accomplish more accurate channel estimation than that of CSI. To elaborate further, the noise term $w_{k,n}$ dominates \eqref{eq:evs_element}, making the raw EVS difficult to interpret. To address this issue, we have defined a calibration method to reduce the noise effect in the raw EVS, which is formulated as
\begin{equation}
    \label{eq:calibration}
    \varepsilon_k={\frac{\bar{\varepsilon}_{R_k}}{2^{\gamma}}}+
    {\frac{2^{\gamma}-1}{2^{\gamma}}}\cdot\left(\frac{1}{T}\sum_{t=1}^{T}\bar{\varepsilon}_{R_k}(t)\right),
\end{equation}
where $T$ is the number of packets and $\gamma$ is the calibration parameter. Note that the second term dominates with the increment of $\gamma$, causing EVS $\varepsilon_k$ to approximate the time average of the raw EVS, making the smoothing effect more significant. Finally, we can obtain the magnitude or phase information of the EVS via \eqref{eq:evs_amphs} and utilize them as channel characteristics, respectively. In conclusion, the EVS is not only useful for making the phase information usable but it is also estimated by more symbols than CSI. Therefore, we intend to utilize the EVS as the channel feature in the device-free localization system to distinguish the location of a human in an indoor environment.

\subsection{Deep Learning Classifier}

Please note that due to space limitations, we do not provide a detailed methodology of the basis of deep neural networks in this paper. However, we would like to mention a few key points as follows.
After processing the data, the amplitude or phase information of the EVS is used as input to the error vector based classifier. This classifier consists of two hidden layers of DNN. To address the vanishing gradients problem, we use the \textit{SeLU} activation function instead of the commonly used \textit{ReLU} function, due to its smoother function for differentiability \cite{selu}. The classifier has one output layer with the \textit{Softmax} activation function, which calculates the probability of the human location. The loss function computes the loss between the label $c$ and the predicted label $\hat{c}$ using typical \textit{cross-entropy}. Then, it implements the backpropagation mechanism. After the offline training stage, the well-trained model predicts the final result $c_{pre}$ in the online session.

\section{Performance Evaluation}

\begin{figure}
  \centering
    \begin{subfigure}[b]{0.18\textwidth}
    \includegraphics[width=\textwidth]{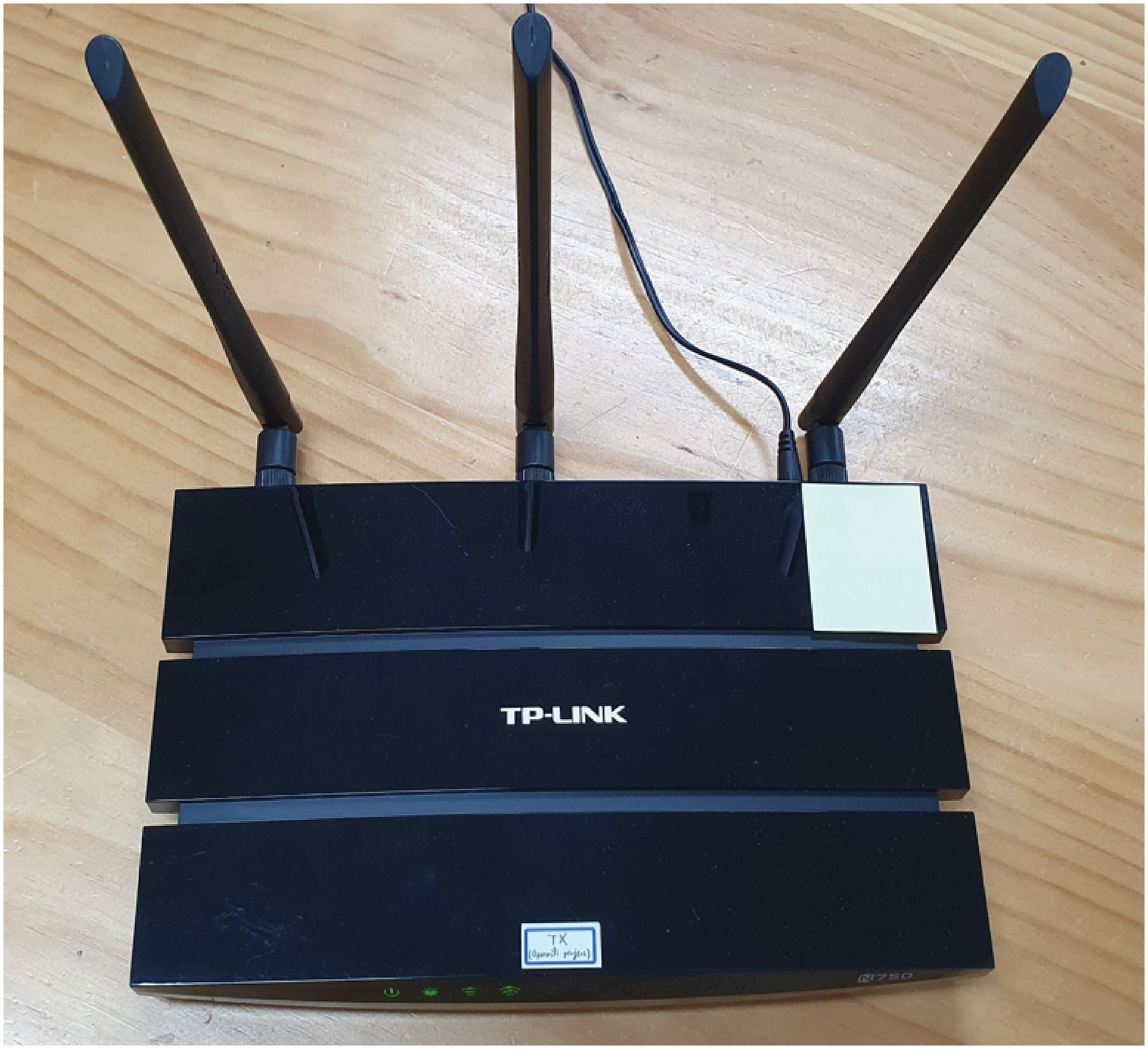}
    \caption{\footnotesize}
    \label{fig:tx}
  \end{subfigure}
  \quad
  \begin{subfigure}[b]{0.25\textwidth}
    \includegraphics[width=\textwidth]{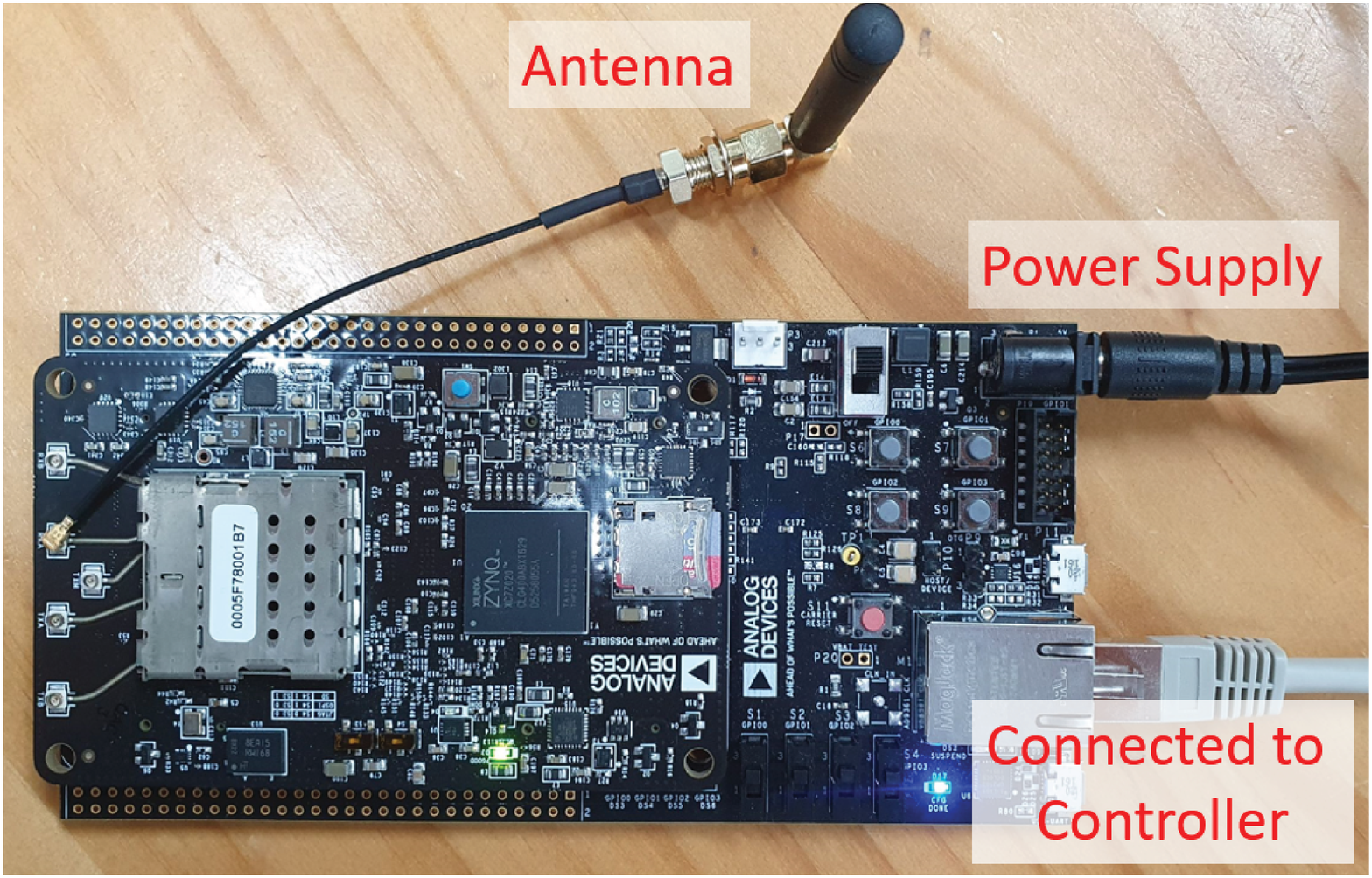}
    \caption{\footnotesize}
    \label{fig:rx}
  \end{subfigure}
  \begin{subfigure}[b]{0.21\textwidth}
    \includegraphics[width=\textwidth]{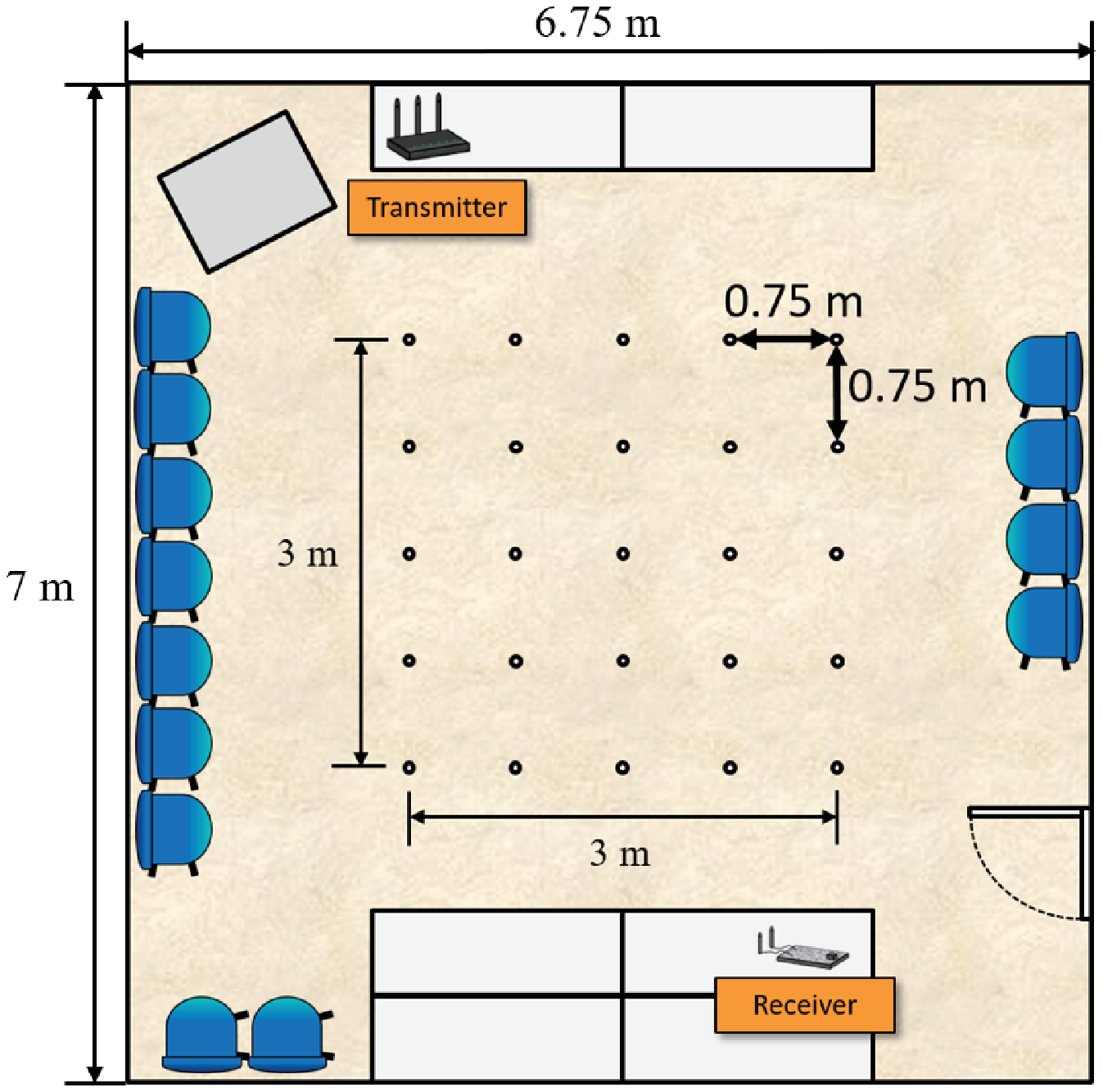}
    \caption{\footnotesize}
    \label{fig:exp_a}
  \end{subfigure}
  \quad
  \begin{subfigure}[b]{0.25\textwidth}
    \includegraphics[width=\textwidth]{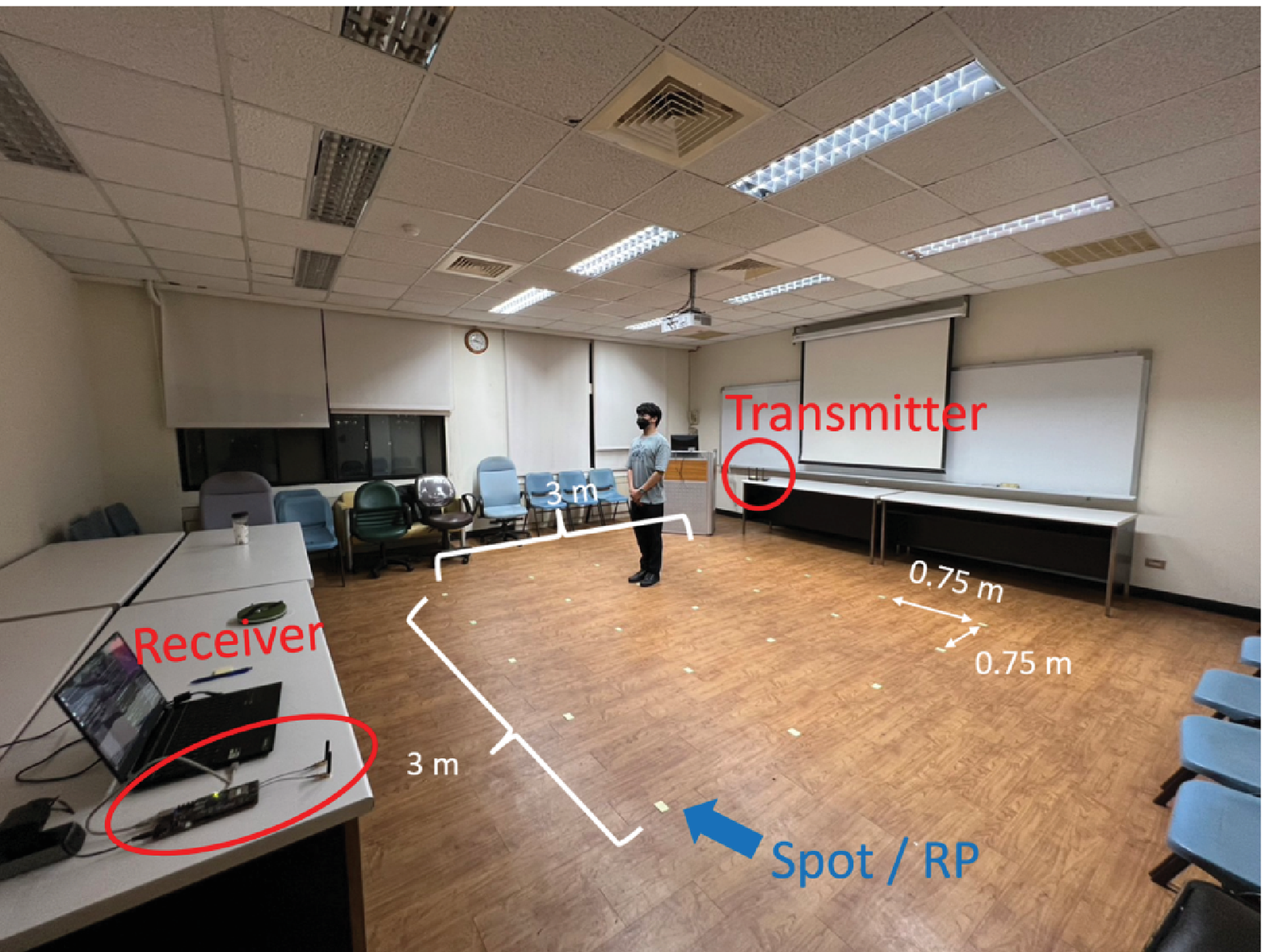}
    \caption{\footnotesize}
    \label{fig:exp_b}
  \end{subfigure}
\caption{Experimental setting includes a (a) commercial transmitter and a (b) receiver operating in OpenWiFi system. (c) Top-view layout. (d) Field trial scene.}
\label{fig:experiment}
\end{figure}

The experimental setup consists of two Wi-Fi routers, one deployed as the transmitter and the other as the receiver. The experiment is conducted in a conference room, as shown in Fig.\ref{fig:experiment}. The transmitter in Fig. \ref{fig:tx} is a commercial Wi-Fi router (TL-WDR 4300) operating at a sample rate of 100 Hz, with an Atheros chip, in access point mode. For the receiver in \ref{fig:rx}, we use a system-on-chip board (Xilinx ADRV9364-Z7020) running the OpenWiFi project \cite{openwifi} in monitor mode. Both routers operate at 5.22 GHz with a bandwidth of 20 MHz, using a single-input single-output Wi-Fi transceiver system equipped with one antenna. The indoor experimental scene, shown in Fig.~\ref{fig:exp_a}, is divided into 25 location spots, with a spacing of $0.75m\times 0.75m$. A person stands still at each spot, as shown in Fig.~\ref{fig:exp_b}. We collect a total of 4000 and 1000 packets for each label as training and testing data, respectively. In total, there are 26 different classification labels, including the empty room case.

\begin{figure}
  \centering
  \begin{subfigure}[b]{0.24\textwidth}
    \includegraphics[width=\textwidth]{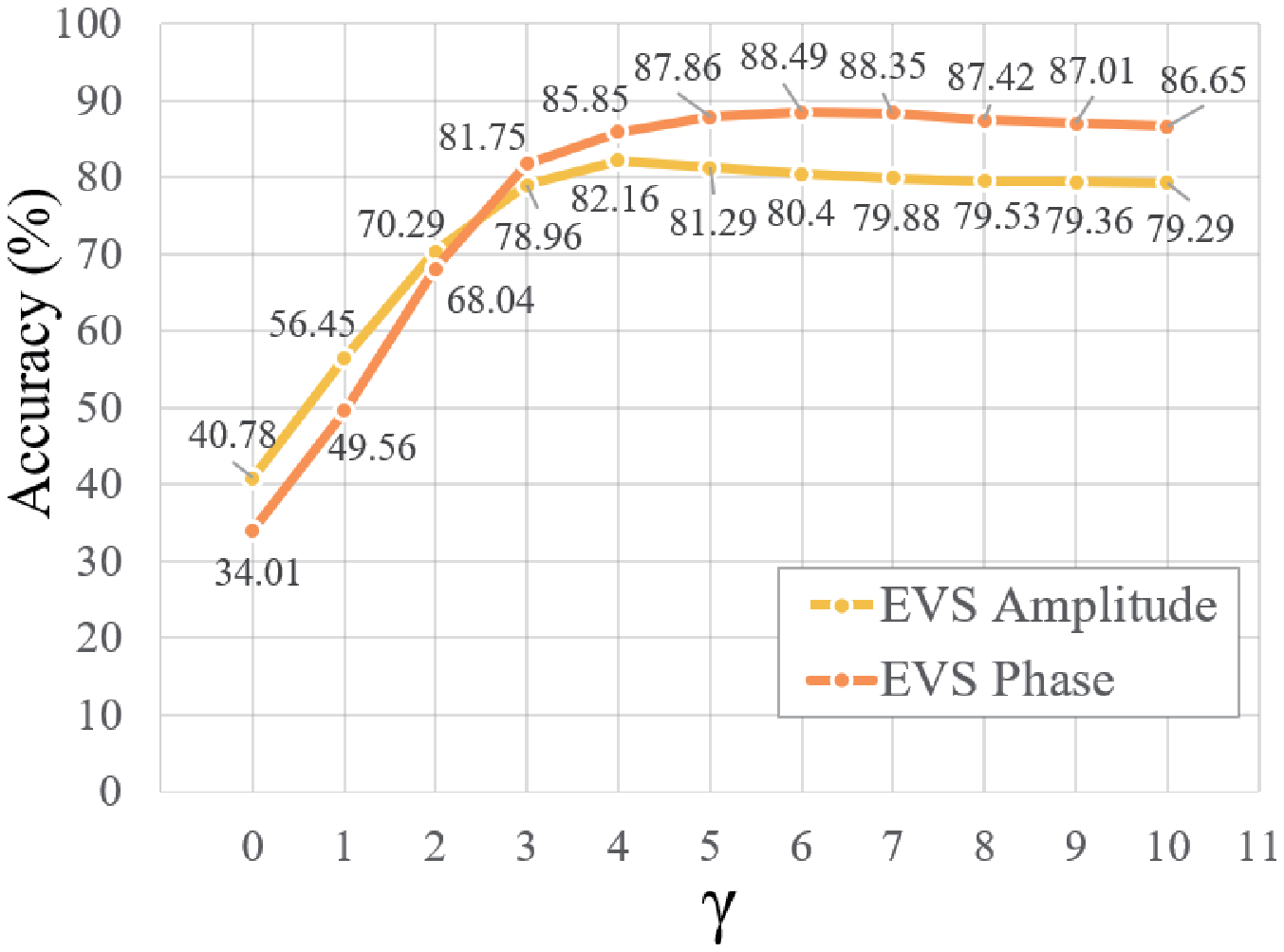}
    \caption{\footnotesize}
    \label{fig:para_a}
  \end{subfigure}
  \begin{subfigure}[b]{0.24\textwidth}
    \includegraphics[width=\textwidth]{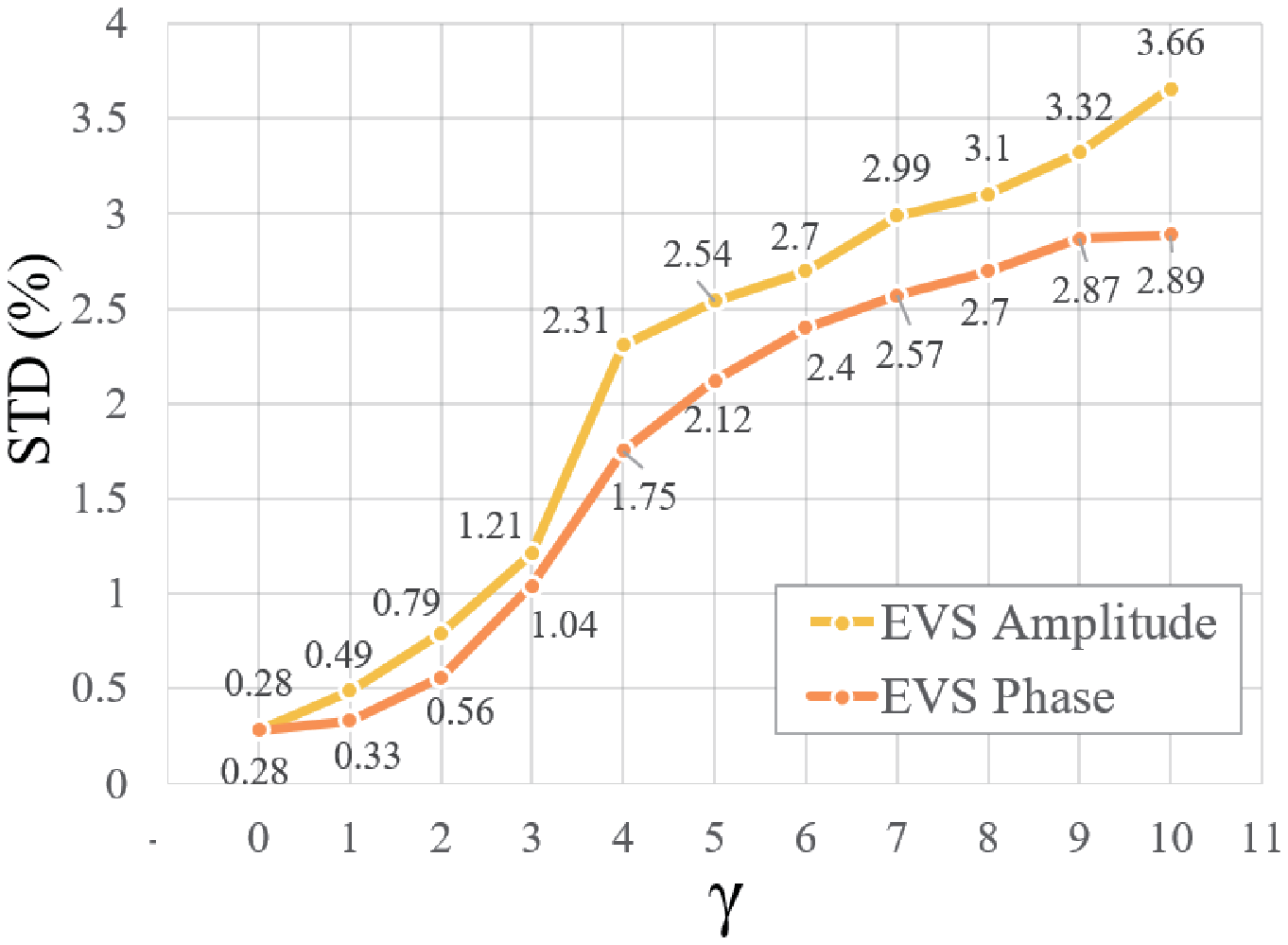}
    \caption{\footnotesize}
    \label{fig:para_b}
  \end{subfigure}
\caption{The performance of EVAL with EVS amplitude and phase under different parameters of $\gamma$ with respect to (a) average accuracy and (b) standard deviation.}
\label{fig:para_comp}
\end{figure}

Fig.~\ref{fig:para_comp} shows the average training accuracy over 50 training runs and its standard deviation (STD) against different calibration parameters $\gamma$. The yellow and tangerine colors represent the amplitude and phase information of the EVS as input training data, respectively. In Fig.~\ref{fig:para_a}, the vertical axis indicates the accuracy. It is observed that larger values of $\gamma$ lead to higher accuracy by effectively smoothing out the AWGN effect in \eqref{eq:evs_element}. The highest accuracy is achieved at $\gamma=4$ and $\gamma=6$ for the EVS amplitude and phase information, respectively. However, calibration does not work when $\gamma=0$ according to \eqref{eq:calibration}. We can also infer that the accuracy of both data types decreases slightly with an increment of $\gamma$ due to excessive smoothing, making the channel feature convergence indistinguishable and eventually leading to an overfitting model during the training phase. This phenomenon is also evident in Fig.~\ref{fig:para_b}, where the STD of accuracy increases with growing $\gamma$. As the amplitude and phase of the EVS can be trained independently, we have the flexibility to set the parameters individually. Based on our observations, we propose using $\gamma$ values of 4 and 6 for the amplitude and phase of the EVS, respectively.

\begin{figure}
  \centering
  \includegraphics[width=0.45\textwidth]{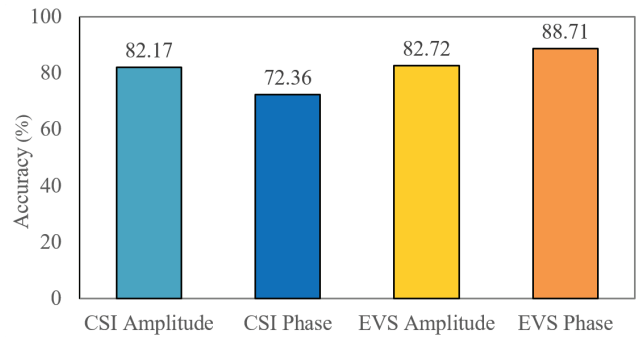}
  \caption{Performance comparison of four different input features of DNN.}
  \label{fig:csievs_comp}
\end{figure}

To verify the performance improvement achieved by EVAL, we collect CSI and I/Q data simultaneously during the experiment. Fig.~\ref{fig:csievs_comp} shows the performance comparison of four different input data types, including CSI amplitude/phase, as well as EVS amplitude/phase. All data types are individually fed into the same DNN learning network. According to Fig.~\ref{fig:csievs_comp}, the EVS phase information outperforms the CSI one, indicating improved classification precision by removing RFO. Furthermore, the amplitude of the CSI in \eqref{eq:csi_element} is the same as that of EVS in \eqref{eq:evs_element} when the exponential term is removed. However, since EVS estimation employs more symbols than that CSI estimation, the EVS amplitude only provides a slight escalation in accuracy compared to that using CSI amplitude. As the phase information provides detail pertinent to the transmission distance, it is potentially more beneficial for indoor localization than amplitude information. Moreover, the EVS phase information provides higher accuracy than the EVS amplitude information.

\begin{figure}
  \centering
  \includegraphics[width=0.45\textwidth]{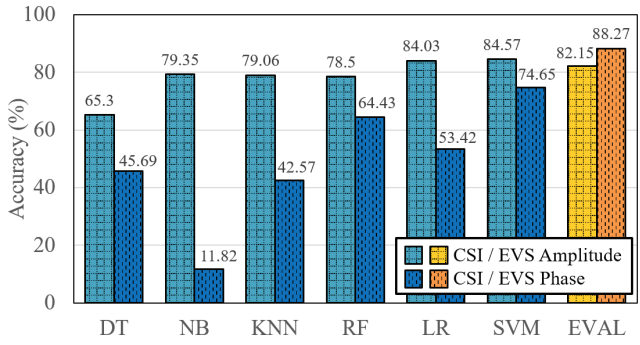}
  \caption{Performance comparison of the proposed EVAL system with other existing benchmarks adopting machine learning methods with the aid of CSI/EVS amplitude/phase data.}
  \label{fig:exter_comp}
\end{figure}

Fig.~\ref{fig:exter_comp} demonstrates the performance comparison of the proposed EVAL system and other existing benchmarks designed based on machine learning methods, including logistic regression (LR) \cite{LR}, decision tree (DT) \cite{DT}, random forest (RF) \cite{CSI_Amp_RF}, support vector machine (SVM) \cite{CSI_Amp_SVM}, k-nearest neighbor (KNN) \cite{CSI_Phs_KNN}, and naive bayes (NB) \cite{CSI_Phs_NB}. The learning models mentioned above are fed with either CSI amplitude information or CSI phase information separately. Fig.~\ref{fig:exter_comp} shows that the classification accuracy of CSI phase information is lower than that of CSI amplitude information, which confirms the presence of the RFO issue. In contrast, our proposed EVAL system not only utilizes more symbols for channel estimation but also eliminates the RFO problem, leading to higher accuracy compared to other existing methods.

\section{Conclusion}
This paper presents a novel method for improving the accuracy of device-free indoor localization using the proposed EVAL scheme in an OpenWiFi system compatible with existing WLANs. The EVS signal, extracted from the I/Q signals, enables the use of rich channel characteristics and eliminates the RFO problem in CSI-based algorithms. Our EVAL system includes a modulation classification and hard-decision decoding method to aid in the extraction of the EVS. Additionally, we propose a calibration method to reduce the effect of AWGN and inject the EVS into our DNN model. Experiments are performed to demonstrate that our proposed EVAL system outperforms conventional machine learning methods and existing device-free indoor localization systems. Overall, our approach provides a more accurate and robust solution for indoor localization, with significant practical implications for a variety of applications.

\linespread{0.95}
\bibliographystyle{IEEEtran}
\bibliography{reference}

\end{document}